\def\finalpaper{1} 

\documentclass[conference]{IEEEtran}
\usepackage{amsmath,amsfonts}
\usepackage{algorithmic}
\usepackage{algorithm}
\usepackage{array}
\usepackage{textcomp}
\usepackage{stfloats}
\usepackage{url}
\usepackage{verbatim}
\usepackage{graphicx}
\usepackage{multirow}
\usepackage{multicol}
\usepackage{cite}
\usepackage{tikz}
\usetikzlibrary{pgfplots.groupplots}
\usepackage{pgfplotstable}
\usepackage{pgfplots}
\usepackage{acronym}
\usepackage{pgf}
\usepackage{datatool}
\usetikzlibrary{positioning}
\usepackage{xfp}
\usepackage{subcaption}
\usepackage{float}
\usepackage{caption} 
\usepackage{circledsteps}
\usepackage{xspace}
\usepackage{tcolorbox}
\usepackage{listings}
\usepackage{siunitx}

\usepackage[acronym,toc]{glossaries}
\newacronym{CDFG}{CDFG}{Control Data Flow Graph}
\newacronym{FSM}{FSM}{Finite State Machine}
\newacronym{QIF}{QIF}{Quantitative Information Flow}
\newacronym{IFA}{IFA}{Information Flow Analysis}
\newacronym{EDA}{EDA}{Electronic Design Automation}
\newacronym{SoC}{SoC}{System-on-Chip}
\newacronym{AST}{AST}{Abstract Syntax Tree}
\newacronym{RTL}{RTL}{Register Transfer Level}

\makeglossaries

\tcbuselibrary{listings}
\newcommand\myCircled[2][]{\ifmmode
\Circled[fill color=black,inner color=white,#1]{\mathsf{#2}}
\else
\Circled[fill color=black,inner color=white,#1]{\sffamily#2}
\fi
}

\usepackage{lipsum}
\AtBeginEnvironment{tabular}{\scriptsize}
\usepackage{background}
\usepackage[usestackEOL]{stackengine}
\setstackgap{L}{\normalbaselineskip}
\SetBgContents{\color{blue}{\tiny \Longstack{PREPRINT - accepted at  IEEE VLSI-DAT, Hsinchu, Taiwan, 2024.}}}
\SetBgPosition{4.5cm,1cm}
\SetBgOpacity{1.0}
\SetBgAngle{0}
\SetBgScale{1.8}

\begin{document}
\bstctlcite{IEEEexample:BSTcontrol}

\title{QTFlow: \underline{Q}uantitative \underline{T}iming-Sensitive Information\\ \centering
\underline{Flow} for Security-Aware Hardware Design on RTL}
\if\finalpaper1

\author{Lennart M. Reimann\IEEEauthorrefmark{1}, Anshul Prashar\IEEEauthorrefmark{1}, Chiara Ghinami\IEEEauthorrefmark{1}, Rebecca Pelke\IEEEauthorrefmark{1},\\ Dominik Sisejkovic\IEEEauthorrefmark{2}, Farhad Merchant\IEEEauthorrefmark{3} and Rainer Leupers\IEEEauthorrefmark{1}\\
\IEEEauthorrefmark{1}RWTH Aachen University, Germany, 
\{lennart.reimann, prashar, ghinami, pelke, leupers\}@ice.rwth-aachen.de\\
\IEEEauthorrefmark{2}Corporate Research, Robert Bosch GmbH, Germany, dominik.sisejkovic@de.bosch.com\\
\IEEEauthorrefmark{3}Newcastle University, farhad.merchant@newcastle.ac.uk \\
\vspace{-1.2cm}
}

\else 
\author{Anonymous Authors \\
Anonymous Authors \\
Anonymous Authors \\
Anonymous Affiliation \\
Anonymous Mails\\
\vspace{-1.2cm}
}
\fi


\maketitle

\begin{abstract}
In contemporary \gls{EDA} tools, security often takes a backseat to the primary goals of power, performance, and area optimization. Commonly, the security analysis is conducted by hand, leading to vulnerabilities in the design remaining unnoticed.  Security-aware EDA tools assist the designer in the identification and removal of security threats while keeping performance and area in mind. Cutting-edge methods employ information flow analysis to identify inadvertent information leaks in design structures. Current information leakage detection methods use quantitative information flow analysis to quantify the leaks. However, handling sequential circuits poses challenges for state-of-the-art techniques due to their time-agnostic nature, overlooking timing channels, and introducing false positives. To address this, we introduce QTFlow, a timing-sensitive framework for quantifying hardware information leakages during the design phase.
Illustrating its effectiveness on open-source benchmarks, QTFlow autonomously identifies timing channels and diminishes all false positives arising from time-agnostic analysis when contrasted with current state-of-the-art techniques. 
\end{abstract}

\glsresetall
\section{Introduction}
In the intricate landscape of modern hardware design, \gls{EDA} has become indispensable due to the increasing design complexity of integrated circuits. These tools adeptly optimize descriptions in terms of both area and performance without compromising functionality. However, most security analyses are conducted manually. The integration of security metrics into \gls{EDA} tools could significantly curtail the incidence of inadvertently implemented and overlooked security vulnerabilities.
Within the domain of \gls{IFA}, a common methodology for establishing security properties like confidentiality~\cite{information_flow_analysis}, the focus is on identifying whether sensitive data can traverse from secure to untrusted hardware components. However, most \gls{IFA} techniques hinge on the non-interference property, which labels any information flow as a threat. Thus, rendering them incapable of distinguishing benign leakages from substantial threats to data security~\cite{non_interference_property}.

In contrast, \gls{QIF} analysis introduces a metric that allows designers to contextualize and prioritize threats~\cite{science_qif}. Current frameworks using \gls{QIF} analysis for hardware lack the ability to consider sequential circuit behavior, leading to increased false positives, especially in area-optimized circuits~\cite{qflow, quardtropy, qflow2}. We address this drawback by introducing a QTFLow to incorporate timing sensitivity into the state-of-the-art framework, called QFlow~\cite{qflow}. Additionally, the methodology allows for the automatic identification of timing channels---vulnerabilities that allow for retrieving sensitive data from the hardware execution time. Therefore, QTFlow is the first framework to \textit{accurately quantify leakages} in sequential circuits and \textit{automatically detect} timing channels.
The major contributions of this paper are:
    (I) The first introduction of timing-sensitivity into quantitative information flow analysis for hardware.
    (II) Removal of false positives during the evaluation.
    (III) Automatic detection of timing channels in a hardware description.


\section{Preliminaries \& Related Work}

\subsection{Threat Model}
Our examination centers on vulnerabilities introduced in the \gls{RTL}-design process, susceptible to exploitation by adversaries after fabrication. In this study, we assume that the attacker can observe outputs and non-secret inputs of selected hardware modules randomly without changing them. The outputs may disclose sensitive data through leakage paths, such as user data or encryption keys. Leakage paths are routes through the hardware that leak data.  Throughout the attack, the adversary possesses complete knowledge of the design structure.



\subsection{Quantitative Information Flow for Hardware}
As mentioned before, \gls{QIF}~\cite{qif_foundations} enhances the expressiveness of \gls{IFA} through quantitative metrics. In contrast to the non-interference property, quantification facilitates the classification of minor information leakages as negligible. Employing information theory, \gls{QIF} quantifies the threat to a secret processed by a system. The probability distribution of inputs and the system's functionality are used to determine the maximum information leakage about the secret to an output. The calculated value quantifies the leakage of the secret bit.

\subsection{Related Work}
Although \gls{QIF} has shown promising results for analyzing hardware, only three frameworks have been developed in recent years. The frameworks aim to detect vulnerabilities that pose a threat to confidentiality, arising from design errors or malicious modifications known as hardware Trojans.
QIF-Verilog~\cite{qif-verilog} generates a timing-independent data flow graph from Verilog descriptions to quantify information flow from a signal marked as sensitive. The framework assesses the uncertainty introduced by operations on the secret before reaching the top module's output, with higher uncertainty indicating increased obfuscation. Despite its utility, QIF-Verilog's reliance on numerous assumptions may lead to overlooked vulnerabilities, as shown in ~\cite{qflow}.

QFlow~\cite{qflow} takes a distinctive approach by incorporating a bitwise analysis and utilizing the Posterior Bayes Vulnerability as a metric, enhancing the quantization process. It's important to note that QFlow initially supported only a limited attack model; however, this limitation has been addressed in the QFlow extension~\cite{qflow2}.
The scope of the threat model is further broadened with the introduction of QuardTropy~\cite{quardtropy}. QuardTropy introduces the innovative 'g-entropy' metric, assessing vulnerability to information leakage in hardware designs. However, none of these frameworks adequately address the analysis of sequential behavior, such as an \gls{FSM} during quantification—a gap that our methodology in QTFlow successfully bridges to reduce the number of false positives.
\begin{figure*}
    \centering
    \includegraphics[width=\textwidth]{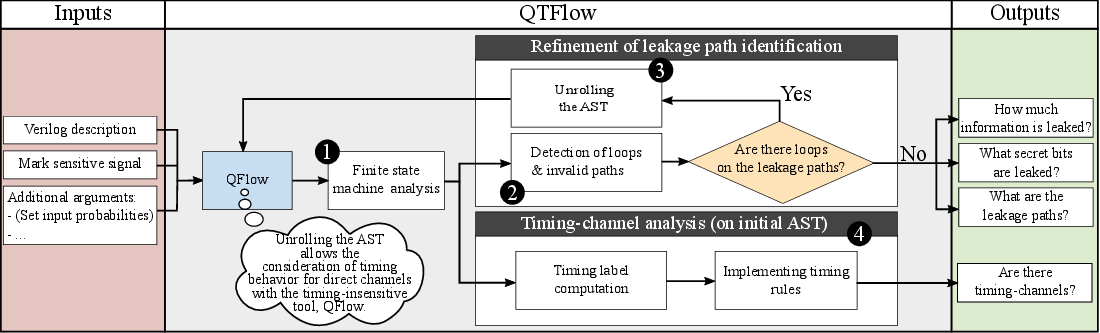}
    \caption{Toolflow of QTFlow.\vspace{-0.4cm}}
    \label{fig:qtflow}
\end{figure*}
\begin{figure}
    \centering
    \includegraphics[width=\columnwidth]{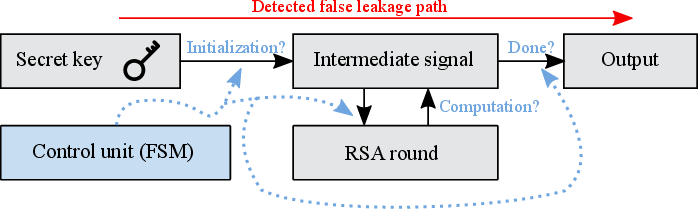}
    \caption{Abstract diagram of an RSA hardware. The Finite State Machine (blue) controls the data flow.\vspace{-0.4cm}}
    \label{fig:rsa_example}
\end{figure}
\section{QTFlow}
QTFlow is constructed based on the QFlow framework, ensuring seamless integration without necessitating any modifications to its existing structure, as illustrated in Fig.~\ref{fig:qtflow}. The methodology is explained using an example of a cryptographic circuit presented in Fig.~\ref{fig:rsa_example} throughout the paper. In the example, the \gls{FSM} regulates the flow of information and computations. It specifically permits the transmission of ciphertext to the output through the "intermediate signal" solely after the completion of the computation. However, without temporal information, as in the case of QFlow, the framework falsely identifies an unauthorized information path from the secret to the output via the intermediate signal, which is infeasible for the actual hardware. The data is processed every RSA round until forwarded to the output, but not prior to that. Thus, a state analysis is required, which derives state transitions for consecutive clock cycles, yielding a raw state sequence that captures the \gls{FSM}'s temporal dynamics. 
\begin{table*}[bp!]
\centering
\vspace{-0.3cm}
\caption{Results of QTFlow on benchmarks with and without Trojans, under different scenarios. Scenario 1: QFlow is executed (time-agnostic), Scenario 2: Only the time-dimension is enabled, Scenario 3: Only timing channel detection is enabled. \label{tab:results} }
\resizebox{\textwidth}{!}{

\begin{tabular}{c|cccc|cccc|cc}
\hline
\multirow{2}{*}{Benchmarks} & \multicolumn{4}{c|}{Scenario 1} & \multicolumn{4}{c|}{Scenario 2} & \multicolumn{2}{c}{Scenario 3} \\ \cline{2-11} 
& \multicolumn{1}{c|}{\begin{tabular}[c]{@{}c@{}}\#Detected/\\ \#Avg. Leakage\end{tabular}} & \multicolumn{1}{c|}{\begin{tabular}[c]{@{}c@{}}\#FP Detected/\\ Avg. Leakage\end{tabular}} & \multicolumn{1}{c|}{\begin{tabular}[c]{@{}c@{}}\#FP Warned/\\ Avg. Leakage\end{tabular}} & \begin{tabular}[c]{@{}c@{}}Time\\ (s)\end{tabular} & \multicolumn{1}{c|}{\begin{tabular}[c]{@{}c@{}}\#Detected/\\ Avg. Leakage\end{tabular}} & \multicolumn{1}{c|}{\begin{tabular}[c]{@{}c@{}}\#FP Detected/\\ Avg. Leakage\end{tabular}} & \multicolumn{1}{c|}{\begin{tabular}[c]{@{}c@{}}\#FP Warned/\\ Avg. Leakage\end{tabular}} & \begin{tabular}[c]{@{}c@{}}Time\\ (s)\end{tabular} & \multicolumn{1}{c|}{\begin{tabular}[c]{@{}c@{}}\#Timing \\ Channels\end{tabular}} & \begin{tabular}[c]{@{}c@{}}Time\\ (s)\end{tabular} \\ \hline
\hline
RSA-T100 & \multicolumn{1}{c|}{33/0.5} & \multicolumn{1}{c|}{1/0.023} & \multicolumn{1}{c|}{1/0.006} & 178 & \multicolumn{1}{c|}{32/0.5} & \multicolumn{1}{c|}{0/-} & \multicolumn{1}{c|}{0/-} & 1294 & \multicolumn{1}{c|}{3} & 185  \\ \hline
RSA-T300 & \multicolumn{1}{c|}{33/0.5} & \multicolumn{1}{c|}{1/0.023} & \multicolumn{1}{c|}{1/0.006} & 176 & \multicolumn{1}{c|}{32/0.5} & \multicolumn{1}{c|}{0/-} & \multicolumn{1}{c|}{0/-} & 1348 & \multicolumn{1}{c|}{3} & 182  \\ \hline
SHA-1 160 & \multicolumn{1}{c|}{3/0.082} & \multicolumn{1}{c|}{3/0.082} & \multicolumn{1}{c|}{2/0.011} & 610 & \multicolumn{1}{c|}{0/-} & \multicolumn{1}{c|}{0/-}  & \multicolumn{1}{c|}{0/-} & 2795 & \multicolumn{1}{c|}{0} & 658    \\ \hline
SHA-2 256 & \multicolumn{1}{c|}{0/-} & \multicolumn{1}{c|}{0/-}  & \multicolumn{1}{c|}{1/0.003} & 418 & \multicolumn{1}{c|}{0/-} & \multicolumn{1}{c|}{0/-}  & \multicolumn{1}{c|}{0/-} & 2398 & \multicolumn{1}{c|}{0} & 435   \\ \hline
SHA-2 384 & \multicolumn{1}{c|}{0/-} & \multicolumn{1}{c|}{0/-}  & \multicolumn{1}{c|}{0/-} & 2004 & \multicolumn{1}{c|}{0/-} & \multicolumn{1}{c|}{0/-}  & \multicolumn{1}{c|}{0/-} & 2035 & \multicolumn{1}{c|}{0} & 2042   \\ \hline
SHA-2 512 & \multicolumn{1}{c|}{0/-} & \multicolumn{1}{c|}{0/-}  & \multicolumn{1}{c|}{0/-} & 2130 & \multicolumn{1}{c|}{0/-} & \multicolumn{1}{c|}{0/-}  & \multicolumn{1}{c|}{0/-} & 2133 & \multicolumn{1}{c|}{0} & 2147    \\ \hline
RSA-TjFree & \multicolumn{1}{c|}{11/0.058} & \multicolumn{1}{c|}{11/0.058} & \multicolumn{1}{c|}{7/0.085} & 184 & \multicolumn{1}{c|}{0/-} & \multicolumn{1}{c|}{0/-}  & \multicolumn{1}{c|}{0/-} & 826 & \multicolumn{1}{c|}{3} & 215     \\ \hline
\end{tabular}}
\end{table*}
To integrate sequential behavior into QFlow's analysis and remove the falsely identified leakage paths, we developed QTFlow. For this, it is imperative to scrutinize any \gls{FSM} that plays a role in directing the flow of sensitive information. The following paragraphs describe the newly introduced methodologies marked with\\ \myCircled{1}- \myCircled{4}, with a visual representation provided in Fig.~\ref{fig:qtflow}.

\myCircled{1}\textbf{Finite State Machine Analysis:\\}
First, the sensitive signal in the hardware is identified and labeled. QFlow is executed, which yields a list of leakage paths. QTFlow extracts the \gls{FSM} of the hardware to identify sequential behavior that influences QFlow's identified leakage paths. Within QFlow, the hardware is represented in a graph structure, an \gls{AST}, including all operations, signals, assignments, and conditions. 
QTFlow processes this graph and identifies states, which correspond to sequential logic. Additionally, state transitions are extracted, which are represented by if-else or case statements assigning new values, i.e. new states, to the identified sequential logic. In the \gls{AST}, each conditional statement modifying the sensitive signal is determined. All states and transitions are used to identify the entire \gls{FSM} that controls the flow of the sensitive data.
The initial state, often the reset state initializing registers, is identified by determining the assignments caused by the reset signal. The reset state represents the \gls{FSM}'s starting point.

\myCircled{2}\textbf{Detection of Loops and Invalid Paths:\\}
This process involves detecting loops and invalid paths within leakage paths using the \gls{FSM}, derived for an accurate representation of the system's timing behavior in the \gls{AST} fed to QFlow.
QTFlow needs to follow the following instructions to identify the loops and invalid paths. The instructions are further elaborated using the example in Fig.~\ref{fig:rsa_example}.
\begin{enumerate}
    \item Parse the leakage paths detected by QFlow.
    \item Compute the state sequence for all leakage paths.
    \item Designate the state containing the last data transfer of the leakage path (intermediate signal $\longrightarrow$ output) as the leaking clock cycle. Any additional cycles can only reduce the amount of information the output carries about the sensitive data, e.g. the secret key.
    \item Compare the leakage paths state sequence with the possible transitions of the \gls{FSM}.
    \item Determine invalid paths by finding a leakage path's state sequence with an order that does not align with the \gls{FSM}. State sequences with an intermediate state that overwrites the secret data, e.g. the leakage path secret key $\longrightarrow$ intermediate signal $\longrightarrow$ output, also represent invalid paths.
    \item Identify loops in the path (intermediate signal $\longleftrightarrow$ RSA round) and the number of loop iterations. The minimal number of loop iterations can be determined by finding the minimal number of state transitions required to reach the final state of the leakage path.
    
\end{enumerate}

\myCircled{3}\textbf{Unrolling the \gls{AST}:\\}
The identified loops are then further processed. 
The process involves modifying the \gls{AST}, that is being used for QFlows analysis, using QTFlow's derived information about invalid paths and loops \gls{FSM} analysis. 
For this, we need to unroll the loops in the internal graph structure to represent common non-looped data paths.
For loop inclusion in the \gls{AST}, QTFlow introduces new intermediary signals corresponding to signals in the looping path. The new signals mirror the structure of the original design structure, with the source signal replaced by the intermediary signal from the preceding step in the loop. The minimum number of loop iterations was computed during the state analysis. For the example (Fig.~\ref{fig:rsa_example}), QTFlow lays out the minimum number of RSA computations between the secret key and the output, so that no bypassing of it is possible for the quantification of the leakage. Additionally, QTFlow neglects any invalid paths. The unrolled \gls{AST} is fed back into QFlow to rerun it.
This empowers QFlow to conduct \gls{QIF} analysis cognizant of the temporal aspects of the hardware.

\myCircled{4}\textbf{Timing channel Detection:\\}
For information to be leaked via a timing channel, the execution time needs to be dependent on the secret value. This means that an adversary can gather information by measuring the time of the execution. For the example (Fig.~\ref{fig:rsa_example}), a timing channel would be detected, caused by the number of RSA rounds being dependent on the value of the secret key.

At first, QTFlow determines the sequential dependency list. For creating the sequential dependency list, the initial step involves computing a list of signals that rely on the secret. This computation is conducted for a limited number of cycles following the reset, utilizing information extracted from the \gls{FSM}. The sensitive signal taints other signals during every signal assignment, making them sensitive as well. Additionally, the newly sensitized signals can then taint signals in the following cycles, and so on. The cycle count \textit{when} the signals are tainted are stored for later usage. 
Afterward, the analysis verifies whether a signal undergoes modification under a conditional statement that uses a variable listed in the sequential dependency list as the condition. Moreover, it needs to be determined if the signal in the condition is tainted before the signal assignment in the conditional statement occurs. The conditional assignments represent possible timing channels, which implies that the temporal occurrence of an output value assignment is contingent upon sensitive data. Nevertheless, a final examination is undertaken to ascertain whether the same assignment occurs in both the "if" and "else" cases; if this holds, it signifies that no information about the secret data can be deduced from the chip's timing.

\section{Evaluation}
\subsection{Evaluation Setup}
Our evaluation employs open-source benchmarks infected with Trojans to assess the effectiveness of QTFlow. Design descriptions of cryptographic accelerators containing Trojans that leak encryption keys~\cite{trusthub_benchmarks} and Trojan-less cryptographic circuits~\cite{opencores}, including SHA, MD5, DES, and 3DES circuits are evaluated. The designs are common benchmarks for security-aware \gls{EDA} tools~\cite{qflow, qflow2, quardtropy}.

\subsection{Results}
\begin{figure*}[t]
	\centering
	\begin{subfigure}[c]{\textwidth}
		\centering
		\begin{tikzpicture}
		\begin{groupplot}[group style={
       group name=plot,
       group size=2 by 1,
       xlabels at=edge bottom,
       ylabels at=edge left,
       horizontal sep=0pt,
       vertical sep=0pt,
       /pgf/bar width=0.08cm}, ybar,ymode = log, ybar=1.2pt, log origin=infty, axis x line=bottom, axis y line=left, ymin=0.0007, ymax = 1, xtick={0,15,32,47,63},xticklabels={0,{Exponent},0,31}, yminorticks = false, axis line style={-},  ylabel={Leakage (bit)},ylabel style={xshift=-0.1cm},enlarge x limits=0.014, legend style={at={(14.5cm,-0.4cm)}, anchor=south}, width=\textwidth, height=2.5cm, legend columns=2]
		
		\nextgroupplot[xmin=0, xmax= 35, width=0.88\textwidth,xlabel={Secret bits},xlabel style={yshift=0.1cm, xshift=1cm}]
		\addplot[black!50,fill] file{data/RSA-T100_m5_b2.txt};
		\addplot[blue!60,fill] file{data/RSA-T100_m5_b2_timing.txt};
		\legend{QFlow, QTFlow}
		\addplot[red,sharp plot,update limits=true,line width=1pt,] coordinates { (0,0.020263671875) (35.5,0.020263671875) };
		
		\nextgroupplot[xmin=61, xmax= 63.3, axis x discontinuity = crunch, axis y line=none, width=0.2\textwidth, xtick={61.5,63}, xticklabels={{Modulo},31}, axis x line=middle, x axis line style=-]
		\addplot[red,sharp plot,update limits=true,line width=1pt,] coordinates { (61.7,0.020263671875) (63,0.020263671875) };
		\end{groupplot}
		\end{tikzpicture}
        \vspace{-0.2cm}
		\caption{RSA-T100 leakage.\vspace{-0.2cm}}
		\label{fig:rsa_t100}
	\end{subfigure}
	
	\begin{subfigure}[c]{\textwidth}
		\centering
		\begin{tikzpicture}
		\begin{groupplot}[group style={
       group name=plot,
       group size=2 by 1,
       xlabels at=edge bottom,
       ylabels at=edge left,
       horizontal sep=0pt,
       vertical sep=0pt,
       /pgf/bar width=0.08cm}, ybar,ymode = log, ybar=1.2pt, log origin=infty, axis x line=bottom, axis y line=left, ymin=0.0007, ymax = 1, xtick={0,15,32,47,63},xticklabels={0,{Exponent},0,31}, yminorticks = false, axis line style={-},  ylabel={Leakage (bit)},ylabel style={xshift=-0.1cm},enlarge x limits=0.014, legend style={at={(12.5cm,-0.6cm)}, anchor=south}, legend image code/.code={
        \draw [#1, fill] (0cm,-0.1cm) rectangle (0.2cm,0.25cm); }, width=\textwidth, height=2.5cm]
		
		\nextgroupplot[xmin=0, xmax= 35, width=0.88\textwidth,xlabel={Secret bits},xlabel style={yshift=0.1cm, xshift=1cm}]
		\addplot[black!50,fill] file{data/RSA-T300_m5_b2.txt};
		\addplot[blue!60,fill] file{data/RSA-T300_m5_b2_timing.txt};
		\addplot[red,sharp plot,update limits=true,line width=1pt,] coordinates { (0,0.020263671875) (35.5,0.020263671875) };
		
		\nextgroupplot[xmin=61, xmax= 63.3, axis x discontinuity = crunch, axis y line=none, width=0.2\textwidth, xtick={61.5,63}, xticklabels={{Modulo},31}, axis x line=middle, x axis line style=-]
		\addplot[red,sharp plot,update limits=true,line width=1pt,] coordinates { (61.7,0.020263671875) (63,0.020263671875) };
		\end{groupplot}
		\end{tikzpicture}
        \vspace{-0.2cm}
		\caption{RSA-T300 leakage.\vspace{-0.2cm}}
		\label{fig:rsa_t300}
	\end{subfigure}
		
	\begin{subfigure}[c]{\textwidth}
		\centering
		\begin{tikzpicture}
		\begin{groupplot}[group style={
       group name=plot,
       group size=2 by 1,
       xlabels at=edge bottom,
       ylabels at=edge left,
       horizontal sep=0pt,
       vertical sep=0pt,
       /pgf/bar width=0.08cm}, ybar,ymode = log, ybar=1.2pt, log origin=infty, axis x line=bottom, axis y line=left, ymin=0.0007, ymax = 1, xtick={0,15,32,47,63},xticklabels={0,{Exponent},0,31}, yminorticks = false, axis line style={-},  ylabel={Leakage (bit)},ylabel style={xshift=-0.1cm},enlarge x limits=0.014, legend style={at={(12.5cm,-0.6cm)}, anchor=south}, legend image code/.code={
        \draw [#1, fill] (0cm,-0.1cm) rectangle (0.2cm,0.25cm); }, width=\textwidth, height=2.5cm]
		
		\nextgroupplot[xmin=0, xmax= 35, width=0.88\textwidth,xlabel={Secret bits},xlabel style={yshift=0.1cm, xshift=1cm}]
		\addplot[black!50,fill] file{data/RSA-TjFree_m5_b2.txt};
		\addplot[blue!60,fill] file{data/RSA-TjFree_m5_b2_timing.txt};
		\addplot[red,sharp plot,update limits=true,line width=1pt,] coordinates { (0,0.020263671875) (35.5,0.020263671875) };
		
		\nextgroupplot[xmin=61, xmax= 63.3, axis x discontinuity = crunch, axis y line=none, width=0.2\textwidth, xtick={61.5,63}, xticklabels={{Modulo},31}, axis x line=middle, x axis line style=-]
		\addplot[red,sharp plot,update limits=true,line width=1pt,] coordinates { (61.7,0.020263671875) (63,0.020263671875) };
		\end{groupplot}
		\end{tikzpicture}
        \vspace{-0.2cm}
		\caption{RSA-TjFree leakage.\vspace{-0.2cm}}
		\label{fig:rsa_tjfree}
	\end{subfigure}
	
	\begin{subfigure}[c]{\textwidth}
		\centering
		\begin{tikzpicture}
		\begin{axis}[ybar,ymode = log, ybar=1.2pt, log origin=infty, axis x line=bottom, axis y line=left, ymin=0.0007, ymax = 1, xmin=0, xmax=31, xtick={0,15,31},xticklabels={0,{Secret Bits},31}, yminorticks = false, axis line style={-},  ylabel={Leakage (bit)},ylabel style={xshift=-0.6cm},yscale = 0.2, bar width = 0.08cm, enlarge x limits=0.02, legend style={at={(12.5cm,-0.6cm)}, anchor=south}, legend image code/.code={
        \draw [#1, fill] (0cm,-0.1cm) rectangle (0.2cm,0.25cm); }, width=\textwidth, height=6cm]
		
		\addplot[black!50,fill] file{data/SHA-160_m5_b2.txt};
		\addplot[blue!60,fill] file{data/SHA-160_m5_b2_timing.txt};
		\addplot[red,sharp plot,update limits=true,line width=1pt,] coordinates { (0,0.020263671875) (35.5,0.020263671875) };
		\end{axis}
		\end{tikzpicture}
        \vspace{-1cm}
		\caption{SHA-160 leakage.}
		\label{fig:sha_160}
	\end{subfigure}
	
	\caption{Leakage value comparison between QFlow and QTFlow. The horizontal line indicates the detection (\textcolor{red}{red}) threshold.}
	\label{fig:results_timing_comparison}
	\vspace{-0.4cm}
\end{figure*}
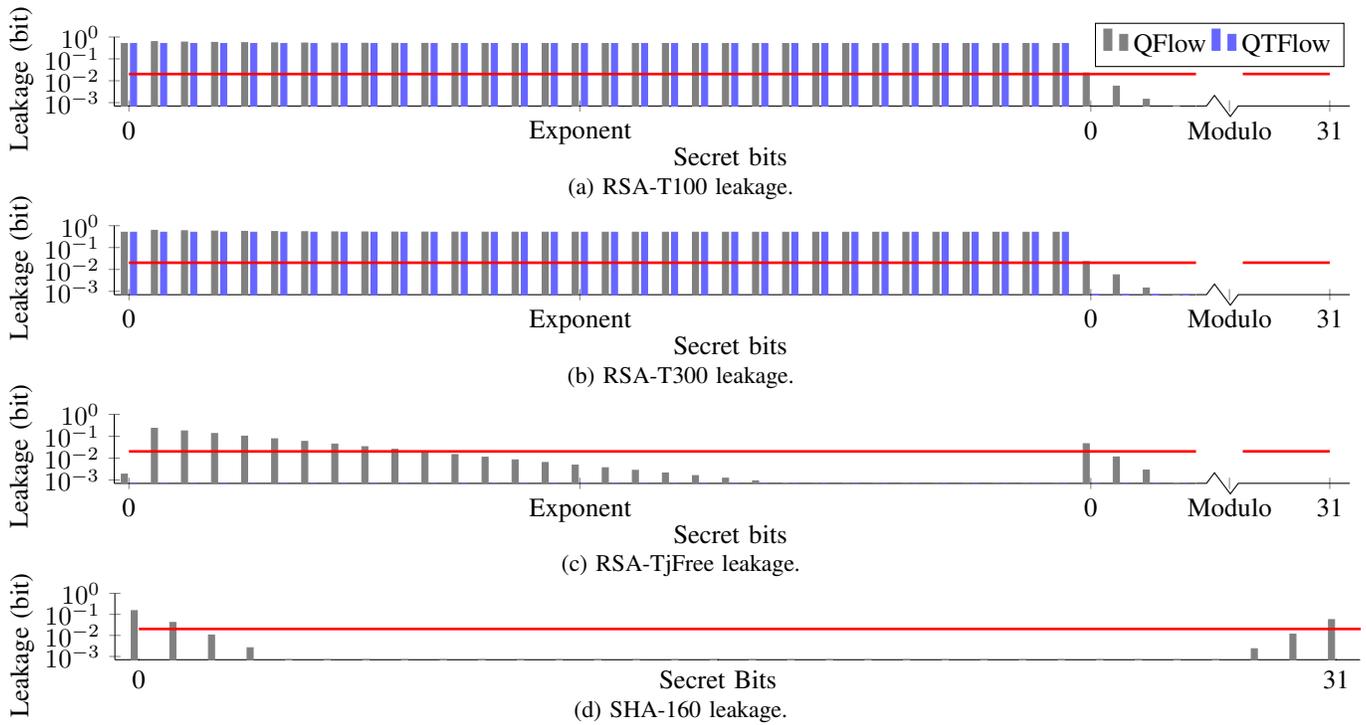
QFlow assesses the likelihood of each secret bit being exposed at the output, assigning a value between \num{0} and \num{1}. A value of 1 indicates direct transmission of the secret to the output, while a diminished value signifies the presence of conditional factors, requiring the attacker to make informed guesses with a computed level of certainty.
The evaluation demonstrated consistent vulnerability detection performance between QTFlow and QFlow across the benchmarks AES, DES, 3DES, and MD5. Their pipelined hardware obviates the necessity of \gls{FSM} involvement in computations. Consequently, the absence of \gls{FSM} utilization eliminates the need to unroll the \gls{AST}. Thus, the results between QFlow and QTFlow are equivalent. However, a slightly higher runtime can be observed, caused by the initial \gls{FSM} analysis in the circuit.
The outcomes for the remaining benchmarks are detailed in Table~\ref{tab:results}. Among the seven benchmarks presented, five exhibit instances of QFlow's false positives (Scenario 1), effectively mitigated through the utilization of our innovative timing-sensitive framework QTFlow (Scenario 2). Notably, the benchmarks featuring false positives demonstrate an increase in analysis time, attributed to the multiple runs of QFlow necessitated by the unrolled \gls{AST}. In the case of the RSA and SHA benchmarks, \textit{QTFlow's timing-sensitivity eliminates all false positives}. Furthermore, the RSA benchmarks illustrate the automatic \textit{detection and identification of timing channels} (Scenario 3), enabling designers to eradicate them and initiate a fresh analysis to verify the successful removal. 
Moreover, Fig.~\ref{fig:results_timing_comparison} presents the changes in computed leakage values for three analyzed benchmarks, comparing QFlow's standalone results with the improved timing-sensitive QTFlow. Among the four benchmarks, only RSA-T100 and T300 have Trojans leaking the exponent part of the secret key.
However, QFlow erroneously identifies data in the RSA-TjFree and SHA-160 benchmarks as leaked, even though no unintentional access to the sensitive information is feasible. Comparable false positives are also detected for the Modulo values in the Trojan-infested RSA benchmarks. \textit{The accurately computed leakages by QTFlow rectify these false positives, enabling precise labeling of vulnerabilities.} No other state-of-the-art \gls{QIF} tools are timing-sensitive, resulting in similar false positives.

\section{Conclusion}
This study introduced timing-sensitivity into a quantitative information flow analysis framework for hardware for the first time. This adaption enhances the security-aware design process at the \gls{RTL}, surpassing the current state of the art, introducing an automatic detection of timing channels, and improving quantification. The efficacy of QTFlow was assessed using open-source hardware benchmarks. Future work can include a combination with formal verification to combine formal assurance with the quantitative metric.
\bibliographystyle{IEEEtran}
\bibliography{bibtexentry}
\end{document}